\documentclass[%
 reprint,
 amsmath,amssymb,
 aps,
prb,
]{revtex4-2}

\usepackage{graphicx}
\usepackage{dcolumn}
\usepackage{bm}

\graphicspath{ {./images/} }
\usepackage{wrapfig}
\usepackage{enumitem}
\usepackage{amsmath,stackengine}
\stackMath
\usepackage{amssymb}
\usepackage{marginnote}
\usepackage{comment}
\usepackage{url}
\usepackage{esint}
\usepackage{multirow}
\usepackage[caption=false]{subfig}
\usepackage{tikz}
\usetikzlibrary{arrows.meta}

\usepackage{mathtools}
\usepackage{listings}
\usepackage[dvipsnames]{xcolor}
\usepackage[colorlinks, allcolors=NavyBlue]{hyperref}
\usepackage{siunitx}
\usepackage{pgfplots}
\pgfplotsset{compat=1.9, , every axis/.append style={font=\scriptsize}}
\usepackage{xspace}
\usepackage{mathrsfs}

\begin{document}

\title{Energy landscape of the kagome antiferromagnet: \\ Characterization of multiple energy scales}

\author{Brandon B. Le}
\email[Corresponding author: ]{sxh3qf@virginia.edu}
\author{Seung-Hun Lee}
\author{Gia-Wei Chern}
\affiliation{
Department of Physics, University of Virginia, Charlottesville, Virginia 22904, USA
}

\date{\today}

\begin{abstract}
We investigate the energy landscape of the kagome Heisenberg antiferromagnet within its coplanar ground-state manifold. Although coplanar states are degenerate at harmonic order, transitions between them require collective weathervane-loop rotations whose barriers grow strongly with loop size. To characterize this structure, we construct disconnectivity graphs using two complementary approaches: exact enumeration and minimax-barrier calculations for small lattices, and a statistical construction for large lattices based on random walks through configuration space, with loop length used as a proxy for barrier height. The exact landscape reveals a dominant low-barrier scale associated with elementary six-spin loops and a broader higher-barrier sector from longer rearrangements. For large systems, the statistical analysis exposes a hierarchy of barrier scales, including a pronounced six-spin-loop peak and an intermediate scale-free regime of loop lengths. This hierarchy provides a natural basis for multiple dynamical time scales: six-spin loops govern the fastest local relaxation, while slower collective dynamics arise from activation of longer loops. These results show that the coplanar manifold is dynamically rugged, with its low-energy dynamics governed by a hierarchy of loop-mediated barriers.
\end{abstract}

\maketitle


\newpage

\section{Introduction}

Geometrical frustration can generate extensively degenerate low-energy manifolds, suppress conventional long-range order, and produce unusual collective and dynamical behavior~\cite{wannier1950antiferromagnetism,Anderson56,pauling1935structure,Lacroix2011,bramwell2001spin,Nisoli13,Udagawa2021,Ramirez1994,Moessner98,Henley2010CoulombPhase,Balents2010}. The kagome Heisenberg antiferromagnet is one of the central examples of this physics and has played a foundational role in the modern understanding of frustration: because of its corner-sharing triangular geometry, the classical nearest-neighbor model admits an extensively degenerate set of ground states satisfying the local $120^\circ$ constraint on every triangle~\cite{zeng1990numerical,Huse1992,harris1992possible,chalker1992hidden}. Thermal and quantum fluctuations partially lift this degeneracy through order-by-disorder effects, selecting coplanar states from the broader classical manifold~\cite{harris1992possible,chalker1992hidden,Reimers1993,chubukov1992order,chubukov1993order,Zhitomirsky2008,taillefumier2014semiclassical,Chern2013,chernyshev2014quantum}. In this sense, the kagome antiferromagnet provides a paradigmatic setting in which local constraints generate a large manifold of low-energy configurations and collective rearrangements, while fluctuations select a restricted but still highly nontrivial low-energy sector.

Even after this selection, however, the coplanar sector remains macroscopically large. Distinct coplanar states are connected by collective weathervane-loop rotations spanning a broad range of lengths and geometries~\cite{Cepas2011,Zhitomirsky2008,Cepas2017}. These loop fluctuations control relaxation and generate multiple time scales in kagome systems, with important differences between winding and nonwinding loops under periodic boundary conditions~\cite{taillefumier2014semiclassical,Saha2021}. The coplanar manifold should therefore be viewed not simply as a set of degenerate states, but as an energy landscape. Although the coplanar states themselves remain degenerate at harmonic order, the paths connecting them are not equivalent: elementary six-spin loops provide the most local rearrangements, whereas longer loops produce increasingly collective moves and typically higher barriers~\cite{chalker1992hidden,chubukov1992order,chubukov1993order,von1993spin,chernyshev2014quantum}. This hierarchy of barrier scales is particularly relevant in light of earlier work pointing to slow, heterogeneous, and glassy-like dynamics in kagome antiferromagnets and related constrained frustrated systems~\cite{cepas2012heterogeneous,cepas2014multiple,taillefumier2014semiclassical}. It is also closely connected to the fact that the coplanar manifold is itself a classical critical state with algebraic correlations: coplanar kagome ground states are equivalent to three-colorings of the dual honeycomb lattice, providing a canonical example of constraint-induced criticality emerging from local rules rather than fine-tuned interactions~\cite{Baxter1970,Kondev96,Chakraborty2002,Castelnovo2005,Cepas2017}.

A useful way to formulate this problem is to regard the coplanar manifold as a configuration-space network, in which nodes represent coplanar states and edges represent single-loop moves~\cite{han2009phase,han2010phase,peng2011self,cao2015ground,lee2021frustration,le2026phase}. In this language, one is interested not only in connectivity, but also in the barrier height associated with each allowed transition. To expose this structure, we use disconnectivity graphs, which provide a compact representation of how configurations cluster into basins and what barriers must be crossed to move between them~\cite{Wales1998,Wales2006,Schon2024,Gallina2021,Gallina2023}. This perspective is well-suited to frustrated systems, where local constraints can organize the low-energy manifold in ways that produce dynamical bottlenecks and slow relaxation even in the absence of disorder. It is especially natural for the kagome problem, where the relevant degrees of freedom are not small-amplitude fluctuations around a unique ordered state, but collective loop rearrangements within a large constrained manifold.

\begin{figure*}
    \centering
    \includegraphics[width=\linewidth]{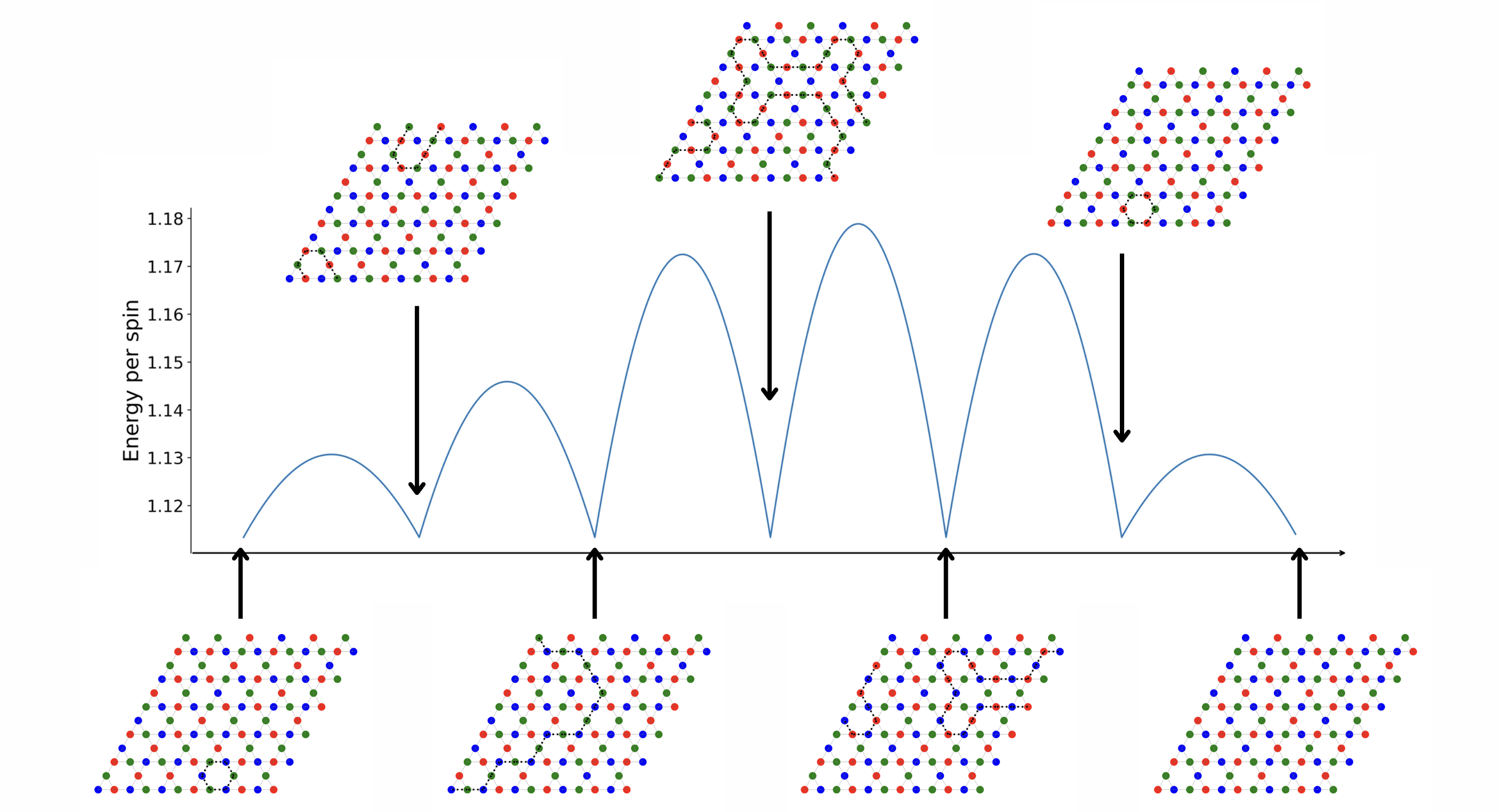}
    \caption{Schematic energy-landscape curve for a $6\times 6$ kagome lattice obtained from successive weathervane-loop rotations between neighboring coplanar ground states. The configurations are shown in the three-state Potts representation, where red, green, and blue denote the three coplanar spin directions separated by $120^\circ$. The dashed loops mark the weathervane loops being rotated. Short, elementary six-spin loops produce the smallest barriers, while longer loops correspond to more collective rearrangements and higher barriers.}
    \label{fig:energy-curve}
\end{figure*}

In this work, we characterize the energy landscape of the coplanar kagome antiferromagnet using exact and statistical disconnectivity graphs. For small lattices, we enumerate the coplanar manifold and compute minimax barriers exactly, while for larger lattices, we sample the manifold by random walks and use loop length as a proxy for barrier height. This reveals a hierarchy of barrier scales, from the elementary six-spin loops that dominate local rearrangements, to a broad scale-free regime of longer collective loops, and finally to winding loops associated with finite-size effects. Our results therefore complement earlier studies of order-by-disorder and loop dynamics by making the landscape structure itself explicit and by identifying how different classes of loop moves organize the connectivity of the coplanar manifold. The rest of the paper is organized as follows: in Sec.~\ref{sec:exact_energy_landscape} we present the exact landscape construction for small systems, in Sec.~\ref{sec:stat-landscape} we introduce the statistical approach for large lattices and analyze the resulting loop-length distributions, and in Sec.~\ref{sec:conclusions} we summarize the main conclusions.

\section{Exact Energy Landscape}
\label{sec:exact_energy_landscape}

The classical nearest-neighbor Heisenberg antiferromagnet 
\begin{eqnarray}
	\mathcal{H} = J \sum_{\langle ij \rangle} \mathbf S_i \cdot \mathbf S_j
    \label{eq:hamiltonian}
\end{eqnarray}
on the kagome lattice is characterized by an extensively degenerate classical ground-state manifold, arising from the local constraint that the three spins on every triangle form a $120^\circ$ configuration. Quantum and thermal fluctuations partially lift this degeneracy and favor coplanar states, in which all spins lie in a common plane while, on each triangle, the three spins still point along three directions separated by $120^\circ$~\cite{chalker1992hidden,chernyshev2014quantum}. Because only these three directions occur in a coplanar ground state, each spin configuration admits a natural representation as a three-coloring, or equivalently as a configuration of the antiferromagnetic three-state Potts model: the three spin directions are labeled by colors such as red, green, and blue, as illustrated in the figure. In this language, the kagome constraint becomes the statement that each triangle must contain all three colors exactly once. Thus, the coplanar kagome ground-state manifold is isomorphic to the ground-state manifold of the antiferromagnetic three-state Potts model.

Throughout this work, the term ``energy landscape of the coplanar manifold'' refers to the barriers along continuous weathervane-rotation paths connecting degenerate coplanar states. For the nearest-neighbor classical Hamiltonian in Eq.~\eqref{eq:hamiltonian}, the classical energy remains unchanged along an ideal weathervane rotation, so the barriers considered in this section are fluctuation-induced barriers obtained from the variation of the harmonic spin-wave zero-point energy along the intermediate, generally noncoplanar configurations. A convenient framework for studying quantum selection within this classically degenerate manifold is linear spin-wave theory. In this approach, spins are mapped to bosonic operators via the Holstein--Primakoff transformation, and the quadratic bosonic Hamiltonian is diagonalized using a real-space Bogoliubov transformation. The resulting magnon energy spectrum determines the zero-point energy of quantum fluctuations, which we average to obtain the energy per spin $E$. It is well established that, to quadratic order in the magnon operators, all coplanar ground states remain degenerate~\cite{chalker1992hidden,chernyshev2014quantum}.

\begin{figure*}[htp!]
    \subfloat[]{\includegraphics[width=0.9\textwidth]{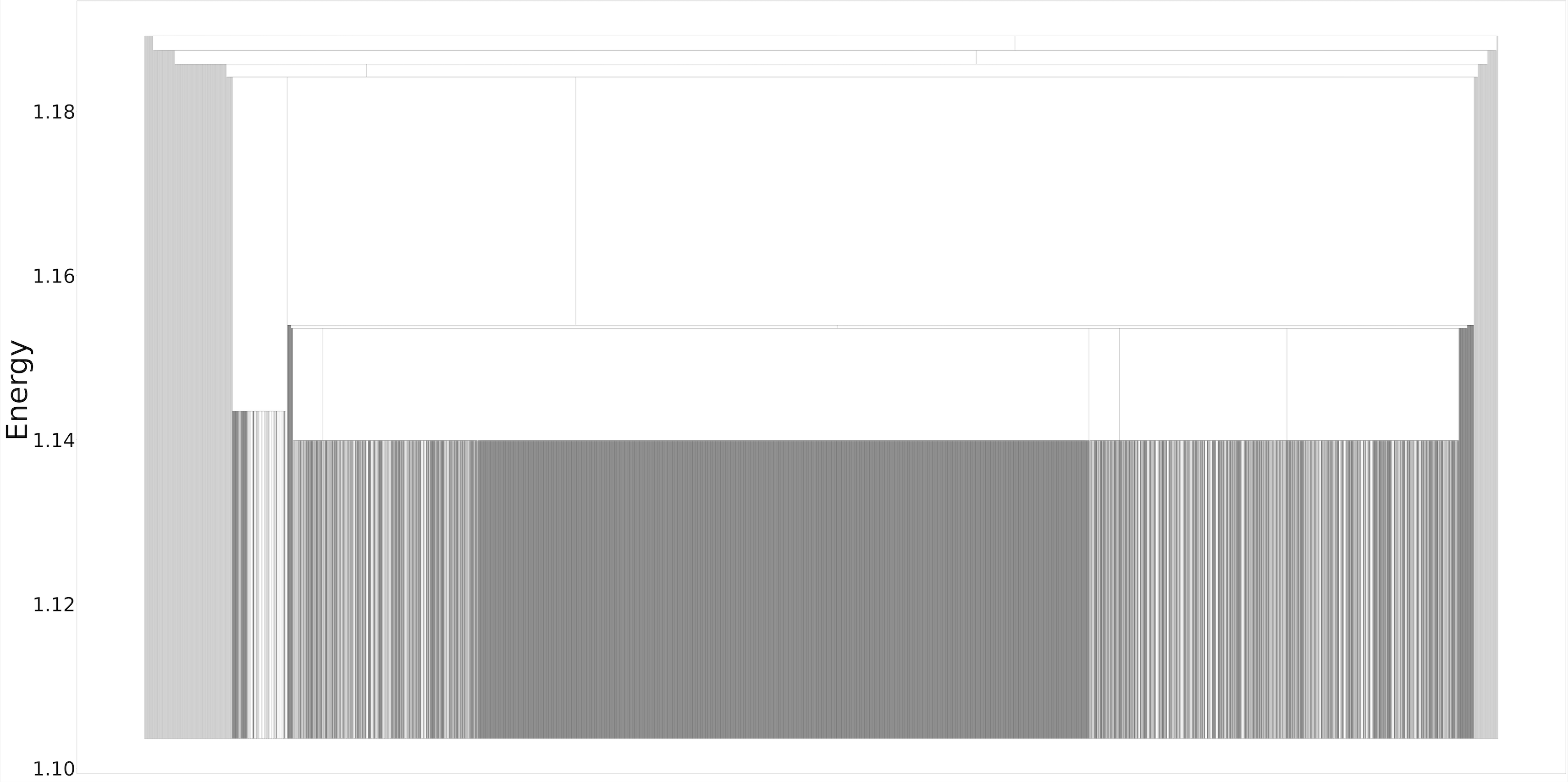}} \\
    \subfloat[]{\includegraphics[width=0.475\textwidth]{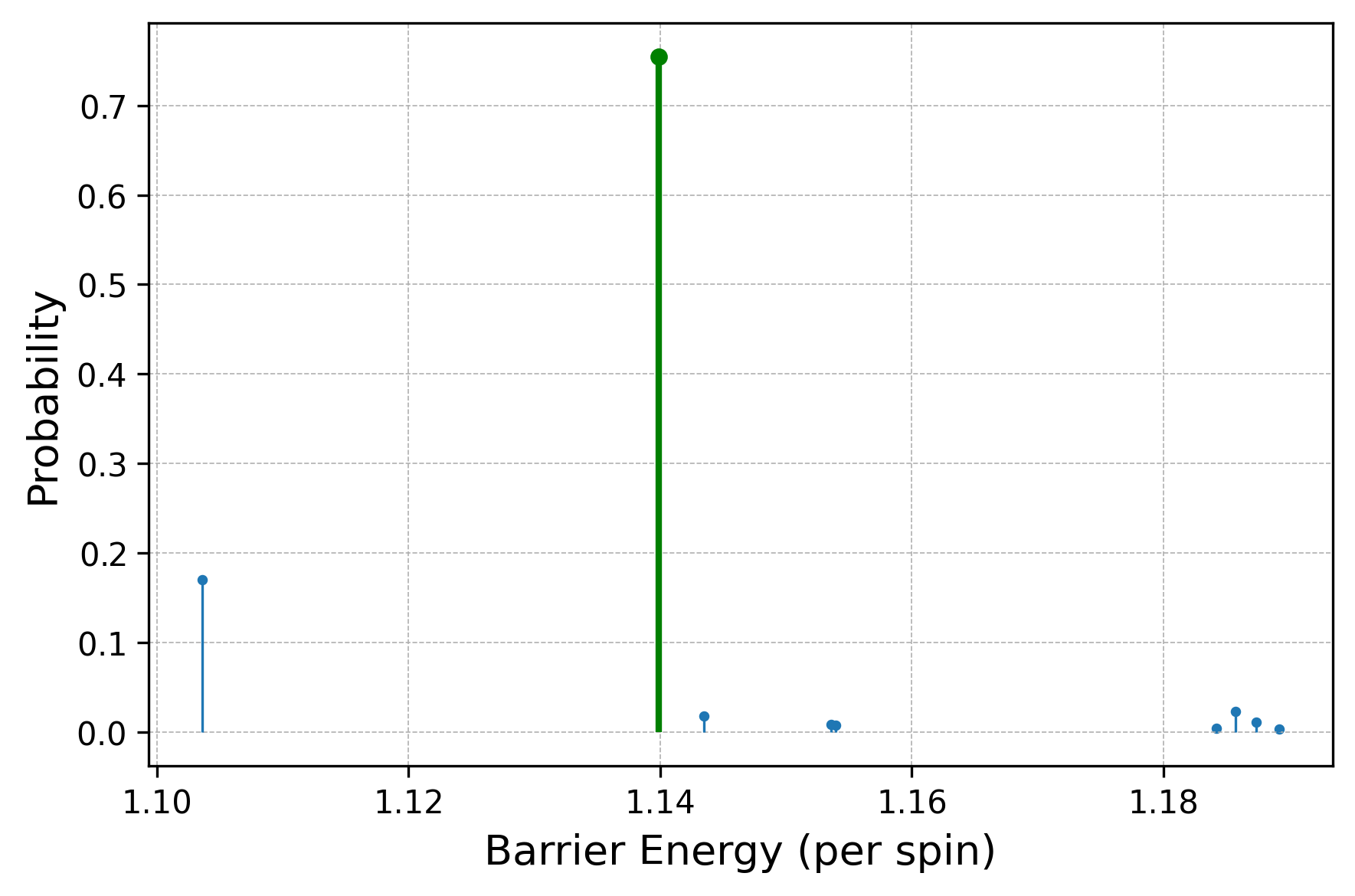}}
    \hfill
    \subfloat[]{\includegraphics[width=0.475\textwidth]{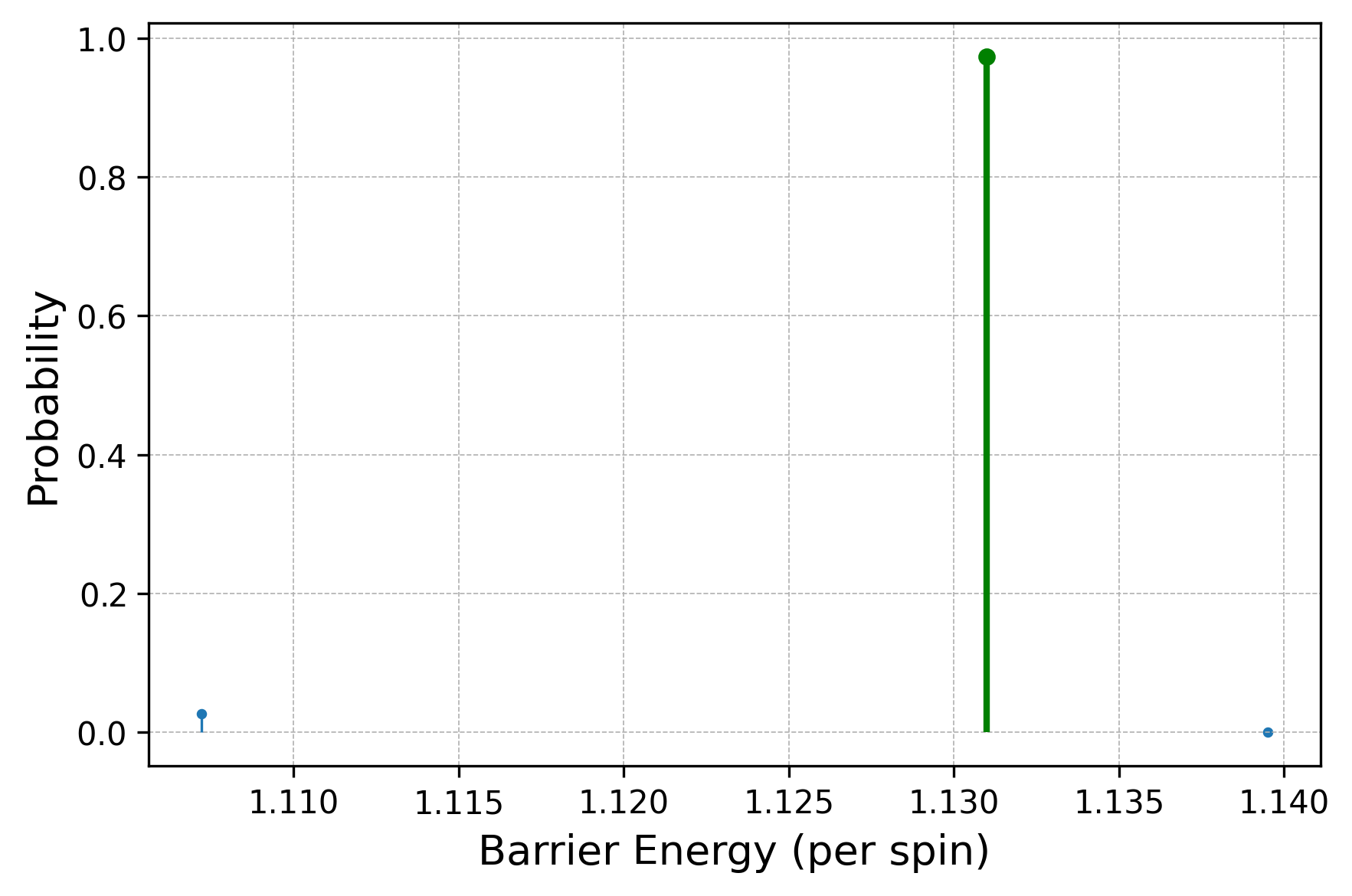}}
    \caption{(a)~Exact disconnectivity graph for a $6\times 3$ kagome lattice. The height at which branches merge gives the minimax barrier energy, namely the largest barrier that must be crossed along the lowest-barrier path connecting two coplanar ground states. (b,c)~Probability distributions of the internal barrier energies of the exact disconnectivity graphs for the $6\times 3$ and $9\times 3$ lattices. In each case, the green dominant peak marks the barrier associated with the elementary six-spin weathervane loop, while the remaining barrier energies are shown in blue.}
    \label{fig:exact_graphs}
\end{figure*}

In Fig.~\ref{fig:energy-curve}, we begin with a random coplanar state on a $6\times 6$ ($N=108$) kagome lattice, represented equivalently as a random three-coloring, and choose a weathervane loop. The figure illustrates one particular sequence of loop rotations rather than all possible routes through the coplanar manifold. In the coloring picture, a weathervane move corresponds to cycling two of the three colors along a closed loop while leaving the third unchanged. We then continuously rotate the spins on that loop from $\theta=0$ to $\theta=\pi$ and compute the corresponding harmonic spin-wave zero-point-energy curve $E(\theta)$. For a fixed initial configuration, selected loop, and prescribed continuous rotation, this energy curve is uniquely determined. Different choices of loops or sequences of loop moves can, however, generate alternative routes through the manifold with comparable or different energy scales. The rotation produces a neighboring coplanar state, after which another available loop is selected and the procedure is repeated. In this way, Fig.~\ref{fig:energy-curve} provides an illustrative picture of the shape and height of the barriers encountered along one route between coplanar ground states. As expected, all coplanar states are degenerate, and the barrier curve between two adjacent coplanar states is symmetric and reaches its maximum at $\theta=\pi/2$. Consistent with previous work~\cite{taillefumier2014semiclassical,Cepas2011}, the typical barrier height increases with the length of the loop.

To visualize the ensemble of such energy barriers, we construct a disconnectivity graph, in which each leaf node corresponds to a coplanar ground state. The height of the smallest barrier connecting two arbitrary coplanar ground state configurations $C_1$ and $C_2$ is the height of the largest energy barrier one must cross to get from $C_1$ to $C_2$. Specifically,
\begin{equation}
    E_{C_1 \to C_2} = \min_{\Gamma \in \mathcal{P}_{C_1 \to C_2}}\max E(\Gamma),
    \label{eq:minmax-E}
\end{equation}
where $\mathcal{P}_{C_1 \to C_2}$ is the set of all paths $\Gamma$ connecting $C_1$ to $C_2$, and $E(\Gamma)$ denotes the set of all of the energies attained in the path $\Gamma$. For example, in Fig.~\ref{fig:energy-curve}, we show one particular $\Gamma$ between the leftmost coplanar configuration $C_1$ and the rightmost coplanar configuration $C_2$. $E(\Gamma)$ is the set of all the energies that appear in the energy curve, and $\max E(\Gamma)$ is the energy value at the highest peak of the energy curve. This definition identifies the lowest possible highest-energy point that must be crossed to transition between the two states. Naturally, in a disconnectivity graph, the vertical axis represents the energy scale so that the relative heights of branching points indicate the energy barriers separating different basins.

For large systems, a complete enumeration of all coplanar configurations is nearly impossible due to exponential growth with system size. Small finite-size lattices, however, allow for a complete characterization of the coplanar ground state manifold and the corresponding energy landscape. The entire coplanar ground state network can be generated by starting from a reference spin configuration, identifying all flippable loops of spins and flipping them, and continuing iteratively. With this characterization, the path $\Gamma$ containing the true energy barrier $E_{C_1 \to C_2}$ between any two arbitrary coplanar configurations $C_1$ and $C_2$ can be efficiently determined using a variation of Dijkstra’s algorithm. Once the minimax barrier for all pairs of configurations is determined, one obtains a symmetric distance matrix representing the relative barrier heights throughout the manifold, which can then be converted into a dendrogram using single-linkage hierarchical clustering. The resulting disconnectivity graph provides a quantitative and visual map of the coplanar manifold, revealing which configurations are closely connected, which require high-energy rearrangements to reach, and how the manifold’s energy landscape is organized.

The exact disconnectivity graph for a $6\times 3$ ($N=54$) kagome system is displayed in Fig.~\ref{fig:exact_graphs}(a). There are 4752 distinct coplanar ground states in total, organized into multiple interconnected basins separated by high-energy barriers. In Fig.~\ref{fig:exact_graphs}(b), we plot the normalized internal-barrier distribution $P(E)$ of the exact $6\times 3$ disconnectivity graph. Specifically, if $N_E$ denotes the number of internal merge barriers with energy $E$, then
\begin{equation}
    P(E) = \frac{N_E}{\sum_{E'}N_{E'}},
\end{equation}
so $P(E)$ is a normalized count over the internal branch heights of the exact disconnectivity graph. By construction, these are the minimal barriers required to bring previously disconnected regions of configuration space into contact. Physically, they are therefore the barriers relevant for global navigation of the landscape: only barriers that lie on some optimal connecting path appear, while higher barriers that can be avoided do not. In this sense, $P(E)$ provides a statistical characterization of the effective barrier structure. Peaks identify barrier heights that recur frequently in optimal paths between minima, while the overall shape reveals the hierarchy and relative abundance of dynamically relevant barriers.

A corresponding distribution is shown in Fig.~\ref{fig:exact_graphs}(c) for the $9\times 3$ ($N=81$) kagome lattice, which contains $261,636$ distinct coplanar states. In both systems, the most prominent feature is the barrier associated with the elementary six-spin (hexagon) weathervane loop, which appears as the dominant peak and represents the basic local rearrangement of the kagome antiferromagnet, for example, between the first and second configurations in Fig.~\ref{fig:energy-curve}. In the $6\times 3$ lattice, this six-spin barrier is already the most probable scale, but it coexists with several additional, less probable barrier energies coming from longer loops. In the $9\times 3$ lattice, by contrast, the distribution becomes much more strongly concentrated on the six-spin barrier, while the remaining barriers carry only very small weight. We note that the six-spin barrier lies at a slightly lower value in the $9\times 3$ system than in the $6\times 3$ system because $E$ is measured per spin.

Beyond this dominant low-energy peak, both distributions also exhibit a second, higher barrier sector associated with longer loops that connect more distant regions of configuration space. In the smaller $6\times 3$ system, this higher-energy structure appears as a handful of discrete subleading peaks, whereas in the larger $9\times 3$ system, it is pushed to very small probability and becomes more weakly resolved. Taken together, these features show that the energy landscape contains two qualitatively distinct kinds of rearrangements: a dense low-barrier sector built from elementary six-spin-loop fluctuations, and a much sparser higher-barrier sector generated by longer, more global loop moves that control the large-scale connectivity of the manifold.

While the exact finite-system distributions already reveal these characteristic barrier scales, they also have inherent limitations. In small lattices, the combinatorics of loops is strongly constrained, and features such as the spurious peak at the minimum energy reflect zero-energy weathervane modes associated with maximum-length ($\ell = 2N/3$) loops that disappear in the thermodynamic limit. More generally, the finite number of configurations and loop types produces discretization effects and artificial degeneracies. To overcome these constraints and extract statistically robust information about the barrier landscape, we therefore turn next to statistical disconnectivity graphs, which sample representative portions of configuration space and allow us to estimate $P(E)$ and its dominant scales for much larger systems where exact enumeration is no longer feasible.

\section{Statistical Landscape and Characterization of Multiple Energy Scales}
\label{sec:stat-landscape}

For large systems, direct evaluation of the spin-wave energy along every loop-rotation path via exact diagonalization becomes computationally infeasible. We therefore use loop length $\ell$ as a proxy for the typical barrier height. The physical basis for this approximation is as follows. As discussed in the previous section, the energy barrier arises from the change in zero-point energy along a weathervane rotation. Each spin on a weathervane loop has essentially the same local coordination: two neighboring spins rotate with it, while its neighbors of the third color remain fixed. Extending the loop therefore repeats approximately the same local arrangement of rotating and fixed spins. Because the exchange interactions are short-ranged, each additional loop segment produces a similar modification of the spin-wave spectrum and hence a similar contribution to the zero-point energy. These contributions are approximately additive when the loop doesn't occupy a large portion of the system, so the total barrier energy therefore grows approximately linearly with $\ell$ in this regime.

Direct support for this approximation was provided by C\'epas and Ralko, who used real-space spin-wave calculations to evaluate the zero-point-energy barriers of weathervane loops sampled from systems containing up to $N=972$ spins~\cite{Cepas2011}. They found that the barrier grows approximately linearly with loop length for short loops, while barriers for longer loops depend increasingly on the surrounding coloring configuration. The use of loop length as an effective barrier variable was subsequently made explicit by C\'epas and Canals, who modeled color exchanges along closed loops as activated processes with a barrier proportional to the loop length~\cite{cepas2012heterogeneous}. These results provide a physical basis for using $\ell$ as an effective barrier variable in large systems. Within the local-loop regime, $\ell$ captures the typical hierarchy of barrier scales while allowing the landscape analysis to be extended beyond systems accessible to direct spin-wave calculations.

For loops involving a substantial fraction of the system, the limitation is more fundamental. As $\ell$ increases, rotating spins no longer form a small rearrangement within a predominantly fixed background, and the barrier becomes more sensitive to the surrounding coloring. Once $\ell>N/3$, more spins of the two participating colors are rotated than remain unrotated, so it is more natural to regard the unrotated spins of the two participating colors as the local disturbance. As $\ell$ approaches its maximum value of $2N/3$, the motion differs from a global rotation only through this shrinking set of unrotated spins, so the barrier turns over at $N/3$ and ultimately vanishes at $\ell=2N/3$. Thus, $\ell$ captures the leading dependence of the typical barrier height on loop size without assigning an exact value of $E$ to every loop. This is sufficient for the large-system analysis, where the objective is to identify the hierarchy of collective transitions rather than calculate each barrier exactly.

Replacing the barrier energy by loop length makes individual transitions inexpensive to evaluate, but the number of coplanar configurations and possible connecting paths still grows exponentially with system size. Exhaustive construction of the configuration network and evaluation of all minimax paths therefore remain infeasible. Instead, we sample representative portions of the landscape using long random walks through the coplanar manifold. To generate a random-walk update, we select a lattice site uniformly at random, then select uniformly one of its neighbors, which carries a different color. These two colors define an alternating path, which is followed until it closes into a valid loop. The two colors are then exchanged along the loop, producing a neighboring coplanar configuration. The resulting sequence of single-loop transitions defines a sampled path $\Gamma$ through configuration space. For two configurations $C_1$ and $C_2$ encountered along a sampled path $\Gamma$, we define the length-based bottleneck
\begin{equation}
    \ell_{C_1\to C_2}^\Gamma = \max\ell(\Gamma_{C_1\to C_2})
\end{equation}
where $\Gamma_{C_1\to C_2}$ is the portion of the sampled path connecting the two configurations and $\ell(\Gamma_{C_1\to C_2})$ denotes the loop lengths sampled along that portion of the path. Since $\Gamma$ is one particular connection path,
\begin{equation}
    \ell_{C_1\to C_2}^\Gamma\geq\ell_{C_1\to C_2}\equiv\min_{\Gamma'\in\mathcal{P}_{C_1\to C_2}}\max \ell(\Gamma').
\end{equation}
The quantity $\ell_{C_1\to C_2}^\Gamma$ is therefore an upper bound on the optimal length-based bottleneck $\ell_{C_1\to C_2}$. We emphasize the important distinction between the exact construction, which minimizes the bottleneck over all possible paths through the complete configuration network, and $\ell^\Gamma_{C_1\to C_2}$, which is determined by one particular segment of a sampled random walk. It follows that $\ell^\Gamma_{C_1\to C_2}$ depends on the sampled trajectory and generally exceeds the true minimax value when a lower-bottleneck alternative path wasn't sampled. The statistical construction should therefore be interpreted as a path-dependent hierarchical representation of the collective-rearrangement bottlenecks encountered during a random walk, not a direct reconstruction of the complete basin structure. 

Using these sampled bottlenecks as input to the same hierarchical procedure used in Sec.~\ref{sec:exact_energy_landscape}, we construct what we refer to as a statistical disconnectivity graph. Its branch heights characterize the loop-length scales encountered in connecting groups of sampled configurations and provide upper bounds on the corresponding optimal length-based bottlenecks. Within the local-loop regime, the ordering of these branch heights corresponds to the expected ordering of typical barrier magnitudes. For longer loops, the graph continues to characterize the scale of collective rearrangements encountered during sampling without assigning a unique value of $E$ to each transition.

\begin{figure*}[t]
    \centering
    \includegraphics[width=0.9\linewidth]{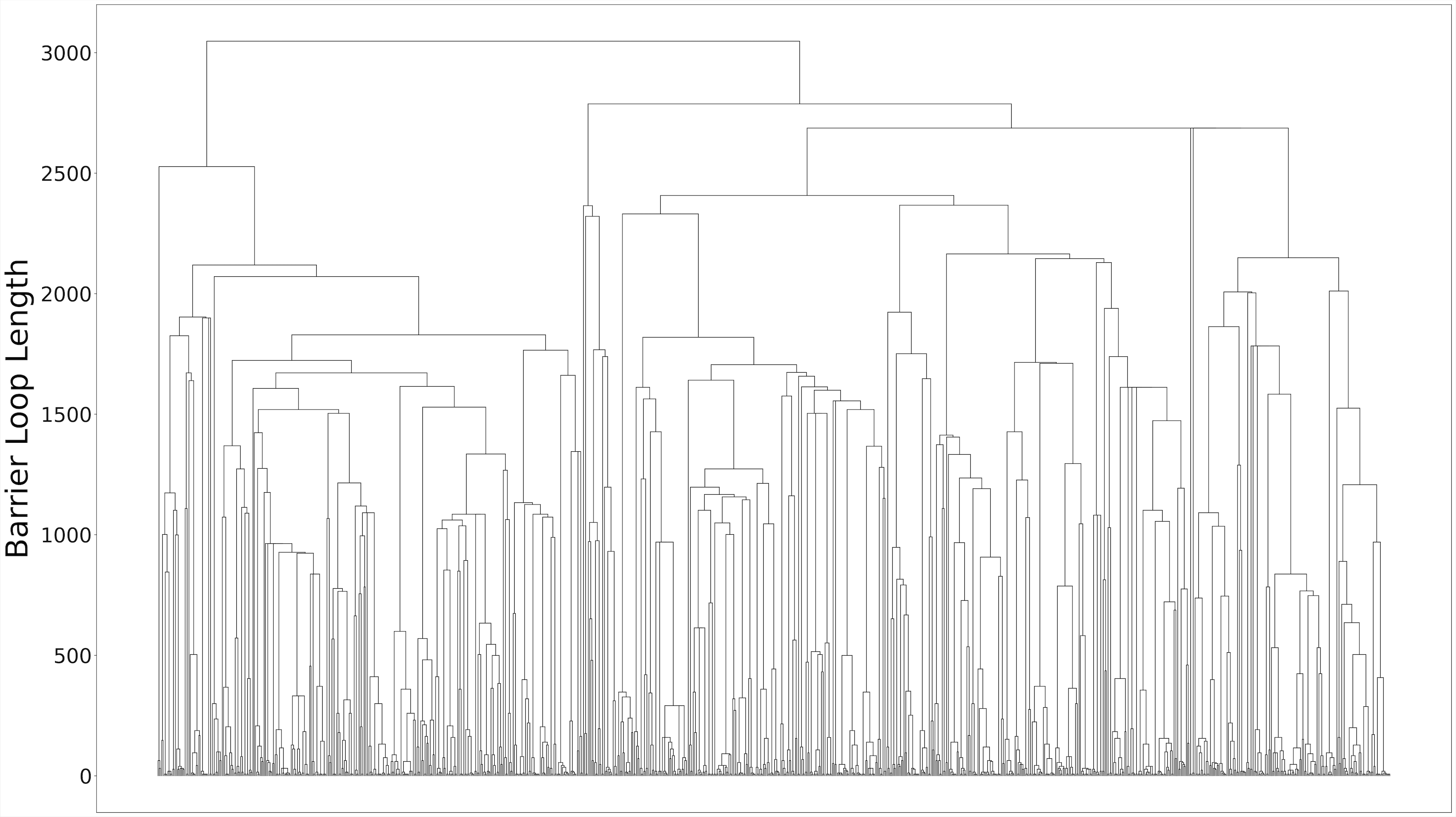}
    \caption{Statistical disconnectivity graph for a $60\times 60$ kagome lattice obtained from a random walk through the coplanar ground-state manifold. Branch heights represent the effective barriers encountered along the sampled path, with loop length used as a proxy for typical barrier energy. The tall, highly branched structure indicates that nearby sampled states are readily connected by local rearrangements, whereas major mergers along the trajectory require much longer collective loops, consistent with a rugged hierarchy of collective-rearrangement scales.}
    \label{fig:L=60_disc}
\end{figure*}

One such disconnectivity graph is displayed in Fig.~\ref{fig:L=60_disc} for a $60\times 60$ ($N=10800$) kagome lattice. The tall, multiply branched structure indicates that, along the sampled trajectories, many nearby configurations are connected through small local rearrangements, whereas major mergers among groups of sampled configurations occur only at substantially larger bottlenecks. This recurring separation within the sampled configuration space is indicative of a landscape whose large-scale connectivity depends on spin rotations involving hundreds or thousands of sites. All branch heights in Fig.~\ref{fig:L=60_disc} lie below the turning point $\ell=N/3=3600$, so the sampled loops remain below the regime in which the barrier begins to decrease as $\ell$ approaches its maximum value. The long vertical segments before major mergers therefore represent large loop-length bottlenecks and, within the approximation discussed above, larger typical barrier energies. These cooperative rearrangements are consequently expected to slow exploration of configuration space despite the absence of energetic differences among the coplanar states themselves. In this sense, the graph provides evidence for emergent ruggedness: local fluctuations permit easy motion within small neighborhoods of the manifold, but accessing distant regions requires rare, highly coordinated moves.

\begin{figure}
    \centering
    \subfloat[]{\includegraphics[width=\linewidth]{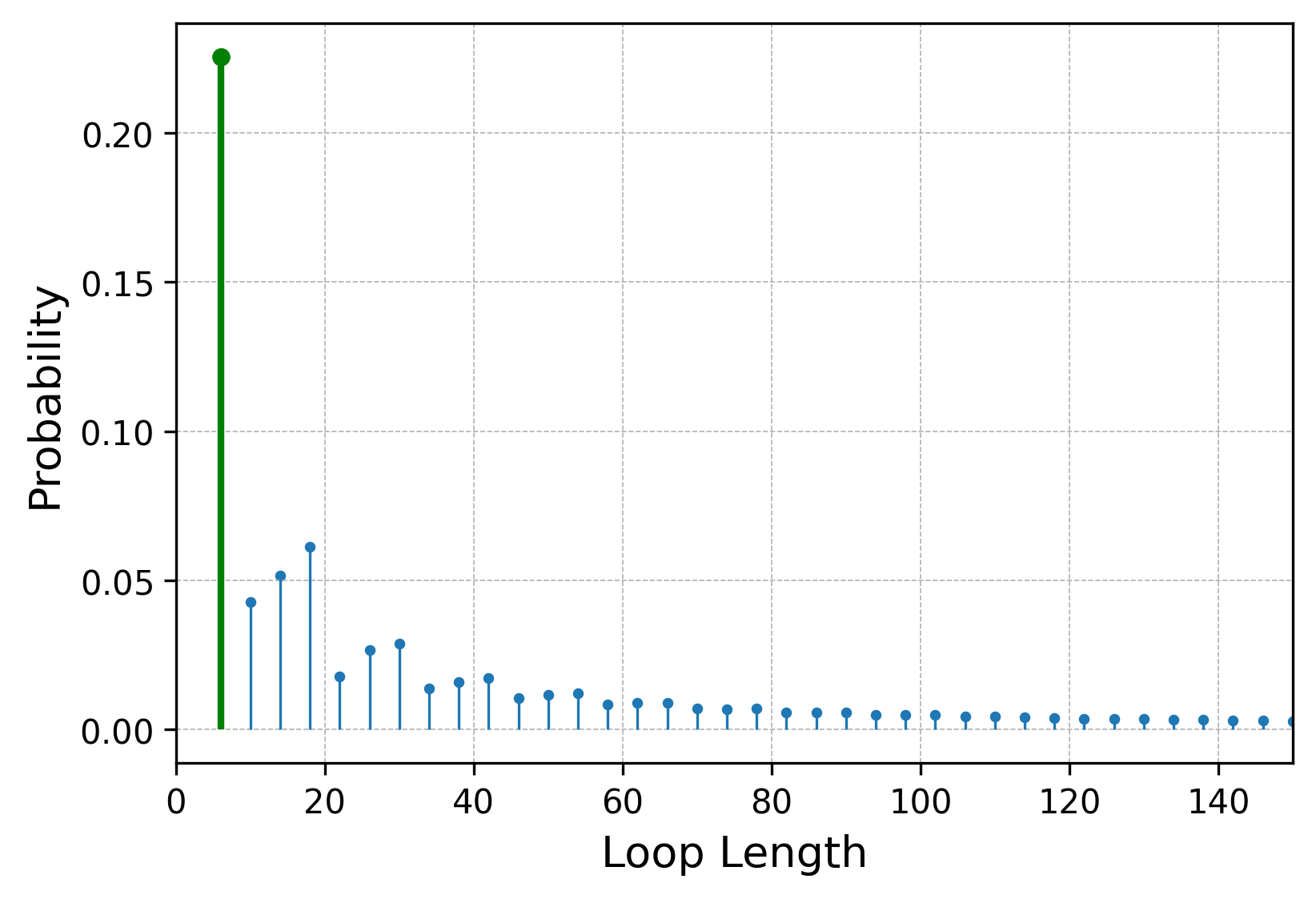}}
    \hfill
    \subfloat[]{\includegraphics[width=\linewidth]{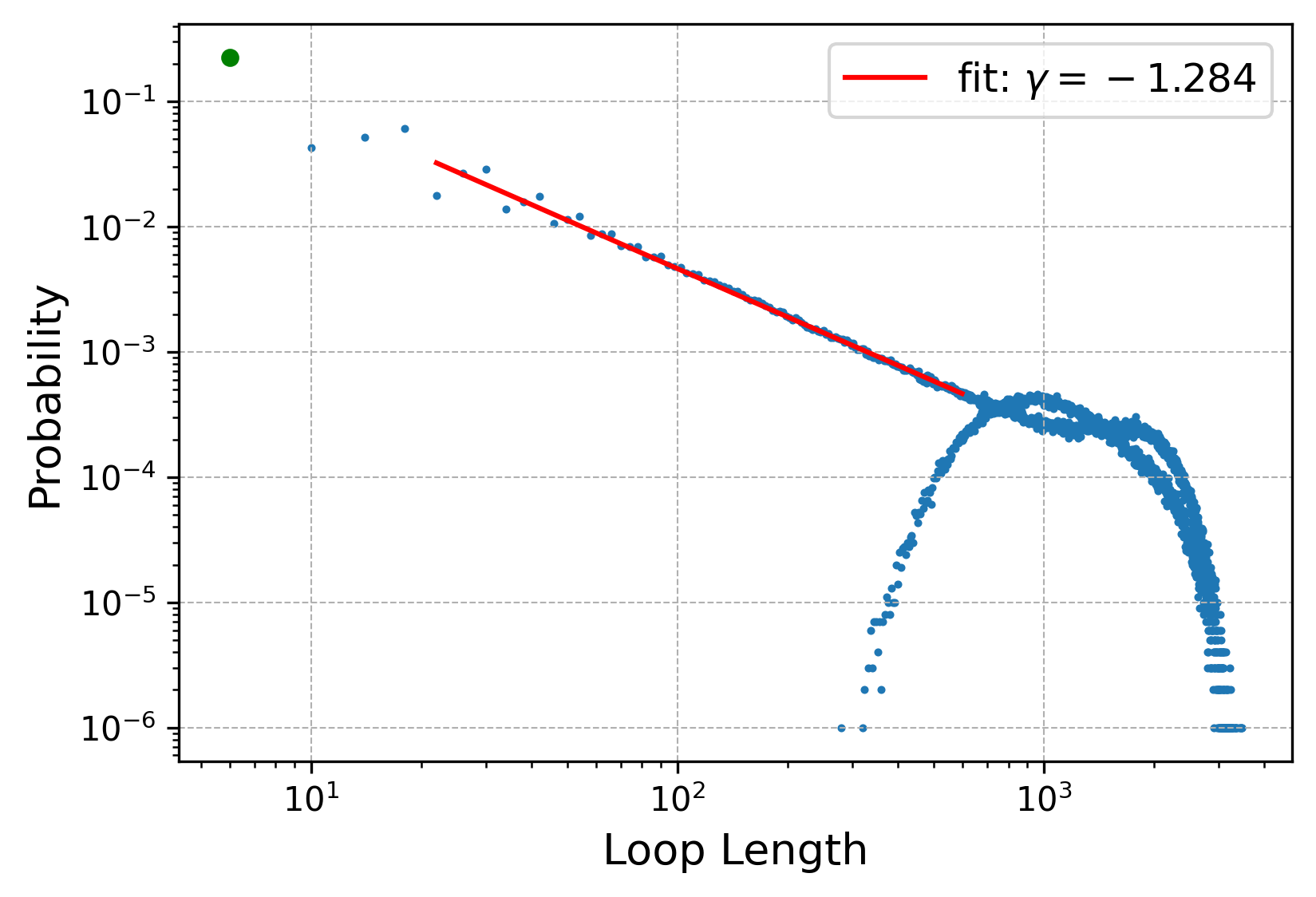}}
    \caption{Distribution of sampled effective barrier scales for a $60\times 60$ kagome lattice, obtained from $10^6$ loop-length measurements aggregated over $100$ independently seeded statistical disconnectivity-graph realizations. (a)~Linear-scale view of the short-loop sector, highlighting the dominant peak at loop length $\ell=6$ from elementary hexagon weathervane loops. (b)~Log-log view of the full distribution, showing the six-spin peak, an intermediate regime with approximate algebraic behavior, and a large-loop sector associated with winding loops and finite-size effects. The red line shows the least-squares fit over $20\leq\ell\leq600$.}
    \label{fig:L=60_loop_length_dist}
\end{figure}

To determine whether the hierarchy visible in the representative graph in Fig.~\ref{fig:L=60_disc} persists across independent trajectories, we analyze the distribution of sampled effective barrier scales over an ensemble of statistical disconnectivity graphs. For the $60\times 60$ lattice, we generated 100 independently seeded random walks. Each realization began from the same reference $\sqrt{3}\times\sqrt{3}$ configuration, and the first $25\,000$ successful loop flips were discarded as an equilibration period. Equilibration was assessed separately by comparing ensembles initialized from the ordered $\sqrt{3}\times\sqrt{3}$ coloring with ensembles initialized from independently pre-randomized configurations. Their loop-length distributions became indistinguishable within the finite-sampling resolution after approximately $5\,000$ loop flips; we used a more conservative equilibration period of $25\,000$ loop flips. The independent random seeds therefore produced distinct randomized configurations within the Kempe sector containing the reference $\sqrt{3}\times\sqrt{3}$ state before measurements began. Accordingly, the distribution $P(\ell)$ discussed below is conditioned on this sector.

For each realization, we recorded $10\,000$ loop length measurements, with successive measurements separated by $1\,000$ successful loop flips. After each such interval, a valid alternating loop was selected using the same procedure described above, and its length was recorded. The intervening flips serve only to advance the random walk and reduce correlations between successive measurements. Each production trajectory therefore contained approximately $10^7$ loop flips, and the complete ensemble contained $10^6$ stored measurements obtained over approximately $10^9$ production loop flips. The principal features were insensitive to substantial reductions in the sampling volume. Restricting the analysis to 50 realizations or to the first $5\,000$ measurements of each realization left the loop length distribution unchanged within statistical fluctuations, and the distributions obtained from the early, middle, and late portions of the production trajectories were also mutually consistent.

The sampled loop lengths define the internal merge heights of the statistical disconnectivity graphs. Each internal branch height is inherited from the effective barrier of a single sampled loop flip at which two previously separate groups of sampled configurations merge. These heights the same structural role as the merge energies $E$ in the exact disconnectivity graph but remain path-dependent proxies rather than exact minimax barriers. If $N_\ell$ denotes the number of sampled internal branch heights of loop length $\ell$, we define the normalized statistical loop length distribution
\begin{equation}
    P(\ell) = \frac{N_\ell}{\sum_{\ell'}N_{\ell'}}.
\end{equation}
Unlike the exact distribution $P(E)$, this distribution is a sampling-induced frequency obtained from the ensemble of statistical disconnectivity graphs. Figure~\ref{fig:L=60_loop_length_dist} presents the resulting distribution after aggregation over the full ensemble of $100$ independent realizations. While Fig.~\ref{fig:L=60_disc} shows the hierarchical organization of the sampled barriers along one representative trajectory, Fig.~\ref{fig:L=60_loop_length_dist} characterizes the relative abundance of the different effective barrier scales encountered throughout the ensemble. The linear scale view in Fig.~\ref{fig:L=60_loop_length_dist}(a), analogous to the exact barrier distributions in Figs.~\ref{fig:exact_graphs}(b) and (c), displays a pronounced peak at $\ell = 6$, showing that the elementary hexagon loop remains the dominant effective barrier scale in the large-system statistical construction. Because each sampled bottleneck $\ell^\Gamma_{C_1\to C_2}$ is an upper bound on the corresponding optimal length-based bottleneck, the statistical construction biases branch-height weight towards larger loop lengths relative to an exact minimax construction on the same sampled states. The corresponding exact loop length distribution is therefore expected to be even more strongly concentrated at the smallest barrier scale, so the sampled six-spin peak provides a conservative measure of the dominance of elementary hexagon rearrangements.

Beyond the six-spin peak, the log-log representation in Fig.~\ref{fig:L=60_loop_length_dist}(b) reveals an approximately algebraic regime at intermediate loop lengths. A least-squares fit of $\log_{10}P(\ell)$ vs. $\log_{10}\ell$ over $20\leq\ell\leq 600$, disregarding points in the second branch (discussed below), gives $P(\ell)\sim\ell^\gamma$ with $\gamma = -1.284$ and $R^2 = 0.994$. Bootstrap resampling over the independent random-walk realizations gives the $95\%$ interval $-1.290\leq\gamma\leq-1.279$ for this fixed fitting prescription, although the fitted value also depends systematically on the selected interval. Varying the fitting limits while remaining within the visibly linear intermediate branch gives approximately $-1.31\lesssim\gamma\lesssim-1.27$. The intermediate regime is also stable over different system sizes. Applying the same fitting procedure to $30\times 30$, $60\times 60$, and $120\times 120$ lattices gives exponents in the same narrow range $-1.31\lesssim\gamma\lesssim-1.27$, indicating that the intermediate branch is a robust scale-free regime over the observed range. Unlike the sharply peaked six-spin sector, the intermediate regime contains no single preferred loop length and instead reflects a broad hierarchy of collective rearrangement scales. Importantly, within the observed range, larger loops occur with an algebraic rather than exponential frequency, consistent with the view that long-range collective fluctuations in the coplanar manifold play an important role.

Finally, a third energy scale emerges from a pronounced finite-size effect, visible in the splitting into two branches at intermediate loop lengths in Fig.~\ref{fig:L=60_loop_length_dist}(b). The splitting into two branches reflects the distinction between nonwinding and winding loops under periodic boundary conditions~\cite{taillefumier2014semiclassical,Cepas2017}. Nonwinding-loop flips preserve the winding numbers of a coloring, whereas the winding-loop flips included in our random walks can connect different winding-number sectors. Even when both types of loops are allowed, however, the coloring manifold remains divided into disconnected Kempe sectors~\cite{Cepas2017}. Because all trajectories originate from the $\sqrt{3}\times\sqrt
{3}$ reference state, our sampling remains within the sector containing these states. This sector contains approximately $89\%$ of all colorings for the fully enumerated $6\times 6$ lattice. Moreover, along the commensurate sequence of rhombic periodic clusters with $L=3n$, the fraction of colorings outside the corresponding even class extrapolates to only approximately $4.6\%$~\cite{Cepas2017}. Since the present $60\times 60$ system belongs to this sequence, these results strongly suggest that the sampled sector remains dominant. The large-loop branch in Fig.~\ref{fig:L=60_loop_length_dist}(b) therefore describes winding loops sampled within this sector rather than an average over all disconnected Kempe sectors. More broadly, loop statistics and the appearance of system-spanning loops have long been recognized as central to kagome glassiness and slow dynamics~\cite{Chandra1993,cepas2012heterogeneous,cepas2014multiple}. The first branch, which persists across all $\ell$, contains only nonwinding loops at small $\ell$, constrained by lattice geometry to lengths congruent to 2 modulo 4~\cite{von1993spin}. This branch dominates the statistics where winding loops are forbidden. The second branch appears at intermediate to large $\ell$ and corresponds to winding loops that traverse the periodic boundaries, which can take any even length. For small $\ell$, winding loops are impossible, but as $\ell$ grows, they populate a new set of lengths, giving rise to the dome-like shape of the second branch. At very large $\ell$, nearly all loops are winding, and the two branches converge at the system-size cutoff. As the system size increases, this winding-loop sector shifts toward larger $\ell$, while the intermediate algebraic branch remains present, further distinguishing the finite-size sector from the intermediate hierarchy of nonwinding loops.

Our results provide a natural energy-landscape perspective on the two-time-scale dynamics found by C\'epas and Canals~\cite{cepas2012heterogeneous}. In their loop model, the shortest hexagon flips remain active at relatively short times, while the slower relaxation is associated with the rearrangement of frozen clusters through longer unjamming loops. They identify this as the spontaneous emergence of two time scales, with the intermediate plateau arising because only the shortest loops can still move while the processes needed to reorganize the jammed regions remain blocked. In our language, this connects naturally to the barrier structure of the coplanar manifold: the six-spin weathervane loop is the dominant low-energy scale, while the rearrangements needed to escape local trapping belong to a distinct higher-barrier sector.

At the same time, our disconnectivity-graph analysis shows that the landscape is richer than a simple two-scale picture. The six-spin loop indeed defines the leading low-barrier process, but beyond it, we find a broad hierarchy of larger barriers rather than a single second scale, and in large systems, this extends into a scale-free intermediate regime before the finite-size winding-loop sector appears at the largest scales. Thus, the two time scales can be understood as the first and most visible consequence of the landscape: fast relaxation is controlled by elementary six-spin loops, while slower relaxation begins when the system must cross into the broader sector of longer-loop barriers that reorganize more extended regions of configuration space. Our results therefore connect their heterogeneous and glassy-like dynamics to the energy landscape by showing how the loop dynamics are organized by a hierarchy of barrier scales within the coplanar manifold.

This interpretation also clarifies the robustness and broader physical implications of the landscape. At finite temperature, the relevant landscape is a free-energy landscape, in which thermal and entropic contributions supplement the zero-point-energy barriers considered here. Anharmonic thermal fluctuations also lift the equivalence of the coplanar states by modifying their statistical weights and enhancing correlations toward the $\sqrt{3}\times\sqrt{3}$ sector~\cite{Chern2013}. Nevertheless, semiclassical simulations show that once coplanar correlations are established, relaxation remains controlled by weathervane-loop fluctuations in an entropically induced free-energy landscape~\cite{taillefumier2014semiclassical}. The organization into local and increasingly collective rearrangements is therefore expected to remain qualitatively robust in the low-temperature coplanar regime, although the individual barrier heights and statistical weights are temperature dependent. The same constrained hierarchy provides a natural microscopic basis for disorder-free glassy-like behavior, as loop-dynamics studies have found metastable trapping, dynamical heterogeneity, broken ergodicity, and a growing characteristic rearrangement length toward lower energies~\cite{cepas2012heterogeneous, cepas2014multiple}. Our static landscape supports this mechanism but does not by itself demonstrate a glass transition, which would require dynamical signatures such as aging or nonergodic correlation functions. Whereas finite temperature modifies the barrier heights and statistical weights within the low-temperature coplanar landscape, a magnetic field changes the landscape more fundamentally by altering the ground-state manifold and selecting distinct coplanar and collinear regimes, including the one-third magnetization state~\cite{zhitomirsky2002field}. Successive changes between field-dependent minima with different magnetizations can produce plateaus and jumps, while the barriers separating them govern metastability, hysteresis, and sweep-rate dependence; the observation of a transient one-third plateau only at intermediate field-sweep rates in a molecular kagome antiferromagnet illustrates this dynamical influence~\cite{narumi2004observation}. A quantitative description of a magnetization staircase would, however, require an explicit field-dependent construction of the competing minima and transition paths beyond the zero-field coplanar landscape considered here.

\section{Conclusions}
\label{sec:conclusions}

We have investigated the energy landscape of the kagome antiferromagnet within its coplanar ground-state manifold, with particular emphasis on the hierarchy of barriers that controls motion between classically degenerate states. Using exact enumeration and minimax-barrier calculations for small lattices, together with statistical disconnectivity graphs for large systems, we identified a sharply dominant low-barrier scale associated with elementary six-spin weathervane loops, which provide the most local rearrangements of the manifold. Beyond this lies a broad intermediate sector of longer loops whose distribution is approximately scale-free, reflecting a wide hierarchy of collective moves rather than a single secondary barrier. At still larger lengths, winding loops appear as a distinct sector in finite systems with periodic boundaries. Taken together, these features show that the coplanar manifold is not simply degenerate, but dynamically rugged, with its connectivity organized primarily by the contrast between ubiquitous local six-spin processes and a broad hierarchy of longer collective rearrangements.

This landscape picture also clarifies how multiple dynamical time scales can emerge in kagome systems. The dominant six-spin barrier naturally sets the fastest local relaxation scale, since these loops are the easiest to activate and therefore govern short-time motion within nearby regions of configuration space. Slower relaxation requires crossing into the broader sector of longer-loop barriers, so collective equilibration is delayed and can become strongly heterogeneous. In this sense, our results provide an explicit landscape basis for earlier observations of two-step and glassy-like kagome dynamics: the familiar separation between fast hexagon motion and slower collective rearrangements is the first manifestation of the barrier hierarchy, while the scale-free intermediate regime suggests that the full dynamics is richer than a simple two-scale picture and instead involves a broad spectrum of relaxation times. The winding-loop sector marks the most global processes in finite systems, but the central thermodynamic result is the coexistence of a sharply dominant six-spin scale with an extended hierarchy of nonwinding collective loops. More broadly, the statistical disconnectivity framework introduced here provides a scalable way to quantify how multiscale cooperative excitations structure the low-energy manifold, and it should be useful for connecting microscopic loop dynamics to slow relaxation, heterogeneous freezing, and other emergent nonequilibrium phenomena in frustrated kagome magnets.

\begin{acknowledgments}
This work was supported by the U.S. Department of Energy, Office of Science, Basic Energy Sciences, through DE-SC0026087.
\end{acknowledgments}

\bibliography{refs}

@PREAMBLE{
 "\providecommand{\noopsort}[1]{}" 
 # "\providecommand{\singleletter}[1]{#1}%" 
}

@article{cepas2012heterogeneous,
  title={Heterogeneous freezing in a geometrically frustrated spin model without disorder: {S}pontaneous generation of two time scales},
  author={C\'epas, Olivier and Canals, Benjamin},
  journal={Phys. Rev. B},
  volume={86},
  number={2},
  pages={024434},
  year={2012},
  doi={10.1103/PhysRevB.86.024434}
}

@article{peng2011self,
  title={Self-similarity of phase-space networks of frustrated spin models and lattice gas models},
  author={Peng, Yi and Wang, Feng and Wong, Michael and Han, Yilong},
  journal={Phys. Rev. E},
  volume={84},
  number={5},
  pages={051105},
  year={2011},
  doi={10.1103/PhysRevE.84.051105}
}

@article{han2009phase,
  title={Phase-space networks of geometrically frustrated systems},
  author={Han, Yilong},
  journal={Phys. Rev. E},
  volume={80},
  number={5},
  pages={051102},
  year={2009},
  doi={10.1103/PhysRevE.80.051102}
}

@article{han2010phase,
  title={Phase-space networks of the six-vertex model under different boundary conditions},
  author={Han, Yilong},
  journal={Phys. Rev. E},
  volume={81},
  number={4},
  pages={041118},
  year={2010},
  doi={10.1103/PhysRevE.81.041118}
}

@article{wannier1950antiferromagnetism,
  title={Antiferromagnetism. {T}he Triangular {I}sing Net},
  author={G. H. Wannier},
  journal={Phys. Rev.},
  volume={79},
  number={2},
  pages={357},
  year={1950},
  doi={10.1103/PhysRev.79.357}
}

@article{pauling1935structure,
  title={The structure and entropy of ice and of other crystals with some randomness of atomic arrangement},
  author={Pauling, Linus},
  journal={J. Am. Chem. Soc.},
  volume={57},
  number={12},
  pages={2680--2684},
  year={1935},
  doi={10.1021/ja01315a102}
}

@article{bramwell2001spin,
  title={Spin ice state in frustrated magnetic pyrochlore materials},
  author={Bramwell, Steven T and Gingras, Michel JP},
  journal={Science},
  volume={294},
  number={5546},
  pages={1495--1501},
  year={2001},
  doi={10.1126/science.1064761}
}

@article{chubukov1992order,
  title={Order from disorder in a kagom{\'e} antiferromagnet},
  author={Chubukov, Andrey},
  journal={Phys. Rev. Lett.},
  volume={69},
  number={5},
  pages={832},
  year={1992},
  doi={10.1103/PhysRevLett.69.832}
}

@article{chubukov1993order,
  title={Order from disorder in a kagome antiferromagnet},
  author={Chubukov, Andrey},
  journal={J. Appl. Phys.},
  volume={73},
  number={10},
  pages={5639--5641},
  year={1993},
  doi={10.1063/1.353624}
}

@article{Anderson56,
  title = {Ordering and Antiferromagnetism in Ferrites},
  author = {Anderson, P. W.},
  journal = {Phys. Rev.},
  volume = {102},
  issue = {4},
  pages = {1008--1013},
  numpages = {0},
  year = {1956},
  month = {May},
  publisher = {American Physical Society},
  doi = {10.1103/PhysRev.102.1008},
  url = {https://link.aps.org/doi/10.1103/PhysRev.102.1008}
}

@book{Lacroix2011,
  title        = {Introduction to Frustrated Magnetism: Materials, Experiments, Theory},
  editor       = {Claudine Lacroix and Philippe Mendels and Fr\'ed\'eric Mila},
  series       = {Springer Series in Solid-State Sciences},
  volume       = {164},
  publisher    = {Springer Berlin Heidelberg},
  address      = {Berlin, Heidelberg},
  year         = {2011},
  isbn         = {978-3-642-10588-3},
  doi          = {10.1007/978-3-642-10589-0}
}

@article{Henley2010CoulombPhase,
  author       = {Henley, Christopher L.},
  title        = {The ``{C}oulomb Phase'' in Frustrated Systems},
  journal      = {Annu. Rev. Condens. Matter Phys.},
  volume       = {1},
  number       = {1},
  pages        = {179--210},
  year         = {2010},
  doi          = {10.1146/annurev-conmatphys-070909-104138},
}

@article{Moessner98,
  title = {Properties of a Classical Spin Liquid: The {H}eisenberg Pyrochlore Antiferromagnet},
  author = {Moessner, R. and Chalker, J. T.},
  journal = {Phys. Rev. Lett.},
  volume = {80},
  issue = {13},
  pages = {2929--2932},
  numpages = {0},
  year = {1998},
  month = {Mar},
  publisher = {American Physical Society},
  doi = {10.1103/PhysRevLett.80.2929},
  url = {https://link.aps.org/doi/10.1103/PhysRevLett.80.2929}
}

@article{Ramirez1994,
  author       = {Ramirez, A. P.},
  title        = {Strongly Geometrically Frustrated Magnets},
  journal      = {Annu. Rev. Mater. Res.},
  volume       = {24},
  pages        = {453--480},
  year         = {1994},
  doi          = {10.1146/annurev.ms.24.080194.002321},
}

@article{Nisoli13,
  title = {Colloquium: Artificial spin ice: Designing and imaging magnetic frustration},
  author = {Nisoli, Cristiano and Moessner, Roderich and Schiffer, Peter},
  journal = {Rev. Mod. Phys.},
  volume = {85},
  issue = {4},
  pages = {1473--1490},
  numpages = {0},
  year = {2013},
  month = {Oct},
  publisher = {American Physical Society},
  doi = {10.1103/RevModPhys.85.1473},
  url = {https://link.aps.org/doi/10.1103/RevModPhys.85.1473}
}

@book{Udagawa2021,
  title        = {Spin Ice},
  editor       = {Masafumi Udagawa and Ludovic Jaubert},
  series       = {Springer Series in Solid-State Sciences},
  volume       = {197},
  publisher    = {Springer International Publishing},
  address      = {Cham, Switzerland},
  year         = {2021},
  isbn         = {978-3-030-70858-0},
  doi          = {10.1007/978-3-030-70860-3},
}

@article{Balents2010,
  author       = {Balents, Leon},
  title        = {Spin liquids in frustrated magnets},
  journal      = {Nature},
  volume       = {464},
  number       = {7286},
  pages        = {199--208},
  year         = {2010},
  doi          = {10.1038/nature08917},
}

@article{Chern2013,
  title = {Dipolar Order by Disorder in the Classical {H}eisenberg Antiferromagnet on the Kagome Lattice},
  author = {Chern, Gia-Wei and Moessner, R.},
  journal = {Phys. Rev. Lett.},
  volume = {110},
  issue = {7},
  pages = {077201},
  numpages = {5},
  year = {2013},
  month = {Feb},
  publisher = {American Physical Society},
  doi = {10.1103/PhysRevLett.110.077201},
  url = {https://link.aps.org/doi/10.1103/PhysRevLett.110.077201}
}

@article{Reimers1993,
  title = {Order by disorder in the classical {H}eisenberg kagom\'e antiferromagnet},
  author = {Reimers, Jan N. and Berlinsky, A. J.},
  journal = {Phys. Rev. B},
  volume = {48},
  issue = {13},
  pages = {9539--9554},
  numpages = {0},
  year = {1993},
  month = {Oct},
  publisher = {American Physical Society},
  doi = {10.1103/PhysRevB.48.9539},
  url = {https://link.aps.org/doi/10.1103/PhysRevB.48.9539}
}

@article{Huse1992,
  title = {Classical antiferromagnets on the Kagom\'e lattice},
  author = {Huse, David A. and Rutenberg, Andrew D.},
  journal = {Phys. Rev. B},
  volume = {45},
  issue = {13},
  pages = {7536--7539},
  numpages = {0},
  year = {1992},
  month = {Apr},
  publisher = {American Physical Society},
  doi = {10.1103/PhysRevB.45.7536},
  url = {https://link.aps.org/doi/10.1103/PhysRevB.45.7536}
}

@article{Zhitomirsky2008,
  title = {Octupolar ordering of classical kagome antiferromagnets in two and three dimensions},
  author = {Zhitomirsky, M. E.},
  journal = {Phys. Rev. B},
  volume = {78},
  issue = {9},
  pages = {094423},
  numpages = {12},
  year = {2008},
  month = {Sep},
  publisher = {American Physical Society},
  doi = {10.1103/PhysRevB.78.094423},
  url = {https://link.aps.org/doi/10.1103/PhysRevB.78.094423}
}

@article{Cepas2011,
  title = {Resonating color state and emergent chromodynamics in the kagome antiferromagnet},
  author = {C\'epas, O. and Ralko, A.},
  journal = {Phys. Rev. B},
  volume = {84},
  issue = {2},
  pages = {020413},
  numpages = {4},
  year = {2011},
  month = {Jul},
  publisher = {American Physical Society},
  doi = {10.1103/PhysRevB.84.020413},
  url = {https://link.aps.org/doi/10.1103/PhysRevB.84.020413}
}

@article{Baxter1970,
    author = {Baxter, R. J.},
    title = {Colorings of a Hexagonal Lattice},
    journal = {J. Math. Phys.},
    volume = {11},
    number = {3},
    pages = {784-789},
    year = {1970},
    month = {03},
    issn = {0022-2488},
    doi = {10.1063/1.1665210},
    url = {https://doi.org/10.1063/1.1665210},
}

@article{Castelnovo2005,
  title = {Quantum three-coloring dimer model and the disruptive effect of quantum glassiness on its line of critical points},
  author = {Castelnovo, Claudio and Chamon, Claudio and Mudry, Christopher and Pujol, Pierre},
  journal = {Phys. Rev. B},
  volume = {72},
  issue = {10},
  pages = {104405},
  numpages = {7},
  year = {2005},
  month = {Sep},
  publisher = {American Physical Society},
  doi = {10.1103/PhysRevB.72.104405},
  url = {https://link.aps.org/doi/10.1103/PhysRevB.72.104405}
}

@article{Kondev96,
title = {{K}ac-{M}oody symmetries of critical ground states},
journal = {Nucl. Phys. B},
volume = {464},
number = {3},
pages = {540-575},
year = {1996},
issn = {0550-3213},
doi = {https://doi.org/10.1016/0550-3213(96)00064-8},
url = {https://www.sciencedirect.com/science/article/pii/0550321396000648},
author = {Jané Kondev and Christopher L. Henley}
}

@Article{Chakraborty2002,
author={Chakraborty, B.
and Das, D.
and Kondev, J.},
title={Topological jamming and the glass transition in a frustrated system},
journal={Eur. Phys. J. E},
year={2002},
month={Nov},
day={01},
volume={9},
number={3},
pages={227-232},
issn={1292-8941},
doi={10.1140/epje/i2002-10071-7},
url={https://doi.org/10.1140/epje/i2002-10071-7}
}

@article{Cepas2017,
  title = {Colorings of odd or even chirality on hexagonal lattices},
  author = {C\'epas, O.},
  journal = {Phys. Rev. B},
  volume = {95},
  issue = {6},
  pages = {064405},
  numpages = {16},
  year = {2017},
  month = {Feb},
  publisher = {American Physical Society},
  doi = {10.1103/PhysRevB.95.064405},
  url = {https://link.aps.org/doi/10.1103/PhysRevB.95.064405}
}

@article{chalker1992hidden,
  title={Hidden order in a frustrated system: {P}roperties of the {H}eisenberg {K}agom{\'e} antiferromagnet},
  author={Chalker, John T and Holdsworth, Peter CW and Shender, EF},
  journal={Phys. Rev. Lett.},
  volume={68},
  number={6},
  pages={855},
  year={1992},
  doi={10.1103/PhysRevLett.68.855}
}

@article{harris1992possible,
  title={Possible {N}{\'e}el orderings of the {K}agom{\'e} antiferromagnet},
  author={Harris, A Brooks and Kallin, Catherine and Berlinsky, A John},
  journal={Phys. Rev. B},
  volume={45},
  number={6},
  pages={2899},
  year={1992},
  doi={10.1103/PhysRevB.45.2899}
}

@article{taillefumier2014semiclassical,
  title={Semi-classical spin dynamics of the antiferromagnetic {H}eisenberg model on the kagome lattice},
  author={Taillefumier, Mathieu and Robert, Julien and Henley, Christopher L and Moessner, Roderich and Canals, Benjamin},
  journal={Phys. Rev. B},
  volume={90},
  pages={064419},
  year={2014},
  doi={10.1103/PhysRevB.90.064419}
}

@article{von1993spin,
  title={Spin tunneling in the kagom{\'e} antiferromagnet},
  author={von Delft, Jan and Henley, Christopher L},
  journal={Phys. Rev. B},
  volume={48},
  number={2},
  pages={965},
  year={1993},
  doi={10.1103/PhysRevB.48.965}
}

@article{cao2015ground,
  title={Ground-state phase-space structures of two-dimensional {$\pm J$} spin glasses: {A} network approach},
  author={Cao, Xin and Wang, Feng and Han, Yilong},
  journal={Phys. Rev. E},
  volume={91},
  number={6},
  pages={062135},
  year={2015},
  doi={10.1103/PhysRevE.91.062135}
}

@article{zeng1990numerical,
  title={Numerical studies of antiferromagnetism on a {K}agom{\'e} net},
  author={Zeng, Chen and Elser, Veit},
  journal={Phys. Rev. B},
  volume={42},
  number={13},
  pages={8436},
  year={1990},
  doi={10.1103/PhysRevB.42.8436}
}

@article{chernyshev2014quantum,
  title={Quantum selection of order in an {XXZ} antiferromagnet on a kagome lattice},
  author={Chernyshev, AL and Zhitomirsky, Mike E},
  journal={Phys. Rev. Lett.},
  volume={113},
  number={23},
  pages={237202},
  year={2014},
  doi={10.1103/PhysRevLett.113.237202}
}

@article{cepas2014multiple,
  title={Multiple time scales from hard local constraints: Glassiness without disorder},
  author = {C\'epas, Olivier},
  journal = {Phys. Rev. B},
  volume = {90},
  issue = {6},
  pages = {064404},
  year = {2014},
  doi = {10.1103/PhysRevB.90.064404}
}

@article{Wales1998,
  title={Archetypal energy landscapes},
  author={Wales, David J and Miller, Mark A and Walsh, Tiffany R},
  journal={Nature},
  volume={394},
  number={6695},
  pages={758--760},
  year={1998},
  doi={10.1038/29487}
}

@article{Wales2006,
  title={Potential energy and free energy landscapes},
  author={Wales, David J and Bogdan, Tetyana V},
  journal={J. Phys. Chem. B},
  volume={110},
  number={42},
  pages={20765--20776},
  year={2006},
  doi={10.1021/jp0680544}
}

@article{Schon2024,
  title={Energy landscapes—{P}ast, present, and future: {A} perspective},
  author={Sch{\"o}n, JC},
  journal={J. Chem. Phys.},
  volume={161},
  number={5},
  pages={050901},
  year={2024},
  doi={10.1063/5.0212867}
}

@article{Gallina2021,
  title={Structural disorder and collective behavior of two-dimensional magnetic nanostructures},
  author={Gallina, David and Pastor, GM},
  journal={Nanomaterials},
  volume={11},
  number={6},
  pages={1392},
  year={2021},
  doi={10.3390/nano11061392}
}

@article{Gallina2023,
  title={Interaction-energy landscapes of kagome and honeycomb lattices of dipole-coupled magnetic nanoparticles},
  author={Gallina, David and Pastor, GM},
  journal={Phys. Rev. B},
  volume={108},
  number={22},
  pages={224412},
  year={2023},
  doi={10.1103/PhysRevB.108.224412}
}

@article{Saha2021,
  title={Spin dynamics of the antiferromagnetic Heisenberg model on a kagome bilayer},
  author={Saha, Preetha and Zhang, Depei and Lee, Seung-Hun and Chern, Gia-Wei},
  journal={Phys. Rev. B},
  volume={103},
  number={22},
  pages={224402},
  year={2021},
  doi={10.1103/PhysRevB.103.224402}
}

@article{Chandra1993,
  title={The anisotropic kagome antiferromagnet: a topological spin glass?},
  author={Chandra, P and Coleman, P and Ritchey, I},
  journal={J. Phys. I France},
  volume={3},
  number={2},
  pages={591--610},
  year={1993},
  doi={10.1051/jp1:1993104}
}

@article{lee2021frustration,
  title={Frustration-induced emergent {H}ilbert space fragmentation},
  author={Lee, Kyungmin and Pal, Arijeet and Changlani, Hitesh J},
  journal={Phys. Rev. B},
  volume={103},
  number={23},
  pages={235133},
  year={2021},
  doi={10.1103/PhysRevB.103.235133}
}

@article{le2026phase,
  title={Phase-space networks and connectivity of the kagome antiferromagnet},
  author={Le, Brandon B and Lee, Seung-Hun and Chern, Gia-Wei},
  journal={arXiv:2601.05933},
  year={2026},
  url={https://doi.org/10.48550/arXiv.2601.05933}
}

@article{zhitomirsky2002field,
  title = {Field-Induced Transitions in a Kagom\'e Antiferromagnet},
  author = {Zhitomirsky, M. E.},
  journal = {Phys. Rev. Lett.},
  volume = {88},
  issue = {5},
  pages = {057204},
  numpages = {4},
  year = {2002},
  month = {Jan},
  publisher = {American Physical Society},
  doi = {10.1103/PhysRevLett.88.057204},
  url = {https://link.aps.org/doi/10.1103/PhysRevLett.88.057204}
}

@article{narumi2004observation,
  title={Observation of a transient magnetization plateau in a quantum antiferromagnet on the Kagom{\'e} lattice},
  author={Narumi, Y and Katsumata, K and Honda, Z and Domenge, J-C and Sindzingre, P and Lhuillier, C and Shimaoka, Y and Kobayashi, TC and Kindo, K},
  journal={EPL},
  volume={65},
  number={5},
  pages={705--711},
  year={2004},
  doi={10.1209/epl/i2003-10164-5}
}

\end{document}